\documentclass{article}

\usepackage[dblblindworkshop, final, nonatbib]{neurips_2025}
\usepackage{graphicx}
\usepackage[utf8]{inputenc} 
\usepackage[T1]{fontenc}    
\usepackage{hyperref}       
\usepackage{url}            
\usepackage{booktabs}       
\usepackage{amsfonts}       
\usepackage{nicefrac}       
\usepackage{microtype}      
\usepackage{xcolor}         
\usepackage{xspace}
\usepackage{inconsolata}

\usepackage[backend=biber,style=numeric]{biblatex}
\usepackage{booktabs}

\usepackage{hyperref}
\newcommand{\repoURL}{https://github.com/hyunjun1121/MacroBench}

\newcommand{\macrobench}{\texttt{MacroBench}\xspace}
\newcommand{\selenium}{Python{+}Selenium\xspace}
\newcommand{\gptfourone}{\texttt{GPT-4.1}\xspace}
\newcommand{\gptfouromini}{\texttt{GPT-4o-Mini}\xspace}
\newcommand{\gemini}{\texttt{Gemini-2.5-Pro}\xspace}
\newcommand{\deepseek}{\texttt{DeepSeek-V3.1}\xspace}
\newcommand{\anonrepo}{\url{https://github.com/hyunjun1121/MacroBench}\xspace}
\newcommand{\airbnblike}{\emph{Airbnb-like}\xspace}
\newcommand{\tiktolike}{\emph{TikTok-like}\xspace}
\newcommand{\redditlike}{\emph{Reddit-like}\xspace}
\newcommand{\instagramlike}{\emph{Instagram-like}\xspace}
\newcommand{\facebooklike}{\emph{Facebook-like}\xspace}
\newcommand{\discordlike}{\emph{Discord-like}\xspace}
\newcommand{\threadslike}{\emph{Threads-like}\xspace}
\usepackage{enumitem}
\setlist[itemize]{noitemsep, topsep=0pt, leftmargin=*}
\usepackage{xurl}             
\usepackage{hyphenat}    
\title{MacroBench: A Novel Testbed for Web Automation Scripts via Large Language Models}
\workshoptitle{Lock-LLM Workshop: Prevent Unauthorized Knowledge Use from Large Language Models}

\author{
  Hyunjun Kim\thanks{Equal contribution.} \\
  KAIST\\
  Daejeon, South Korea\\
  \texttt{hyunjun1121@kaist.ac.kr} \\
  \And
  Sejong Kim$^*$\\
  KAIST\\
  Daejeon, South Korea\\
  \texttt{kingsj@kaist.ac.kr} \\
}

\begin{document}

\maketitle

\begin{abstract}
We introduce \macrobench, a code-first benchmark that evaluates whether LLMs can synthesize \emph{reusable} browser-automation programs (macros) from natural-language goals by reading HTML/DOM and emitting \selenium.
\macrobench instantiates seven self-hosted sites—\airbnblike, \tiktolike, \redditlike, \instagramlike, \facebooklike, \discordlike, and \threadslike—covering \textbf{681} tasks across interaction complexity and targeting difficulty.
Our end-to-end protocol validates generated code via static checks, sandboxed execution, and outcome verification (DOM assertions, database snapshots), and includes a safety suite for scraping, spam/abuse, and credential/privacy prompts.
Across \textbf{2{,}636} model--task runs, we observe stratified success: \gptfouromini\ (96.8\%), \gptfourone\ (95.3\%), \gemini\ (89.0\%), \deepseek\ (83.4\%). Models handle simple tasks reliably (91.7\%) but fail on complex workflows (0.0\%), and none meet production-quality coding practices despite functional completion.
We release our complete benchmark pipeline, evaluation framework, and experimental results at \anonrepo to enable reproducible assessment of macro synthesis for web automation.
\end{abstract}

\section{Introduction}
Large Language Models (LLMs) are increasingly used to automate complex workflows that unfold inside a web browser. We study this problem from a \emph{programmatic} perspective: given a natural-language goal, can an LLM synthesize a reusable, rule-based web automation program (a “macro” or macroinstruction \cite{macromining}) using industry-standard tools such as Python and Selenium? These macros execute concrete browser actions—clicking, typing, submitting forms, following links—so that end users can repeatedly and reliably run them to complete tasks.

\paragraph{A visionless, code-centric agent.}
Unlike multimodal web agents grounded in screenshots or pixels \cite{koh2024visualwebarena,xie2024osworld}, our agent operates purely on HTML/DOM plus programmatic browser actions. This code-first lens reduces perception latency and brittleness, exposes semantic structure (element types, IDs, ARIA roles), and mirrors professional practice (e.g., Selenium locators). It extends DOM-based web interaction research in RL and structured policies \cite{shi2017worldofbits,liu2018wge,jia2019domqnet} while remaining compatible with contemporary prompting paradigms \cite{yao2023react,schick2023toolformer,yao2023treeofthoughts}.

\paragraph{A benchmark for macro synthesis.}
\macrobench isolates three competencies required to translate a natural-language goal into an executable macro:
\begin{enumerate}
    \item \textbf{Code interpretation}: recover task-relevant structure from raw HTML (forms, inputs, buttons, links, and attributes such as \texttt{id}, \texttt{class}, \texttt{name}, \texttt{role}, labels, and hierarchy).
    \item \textbf{Code generation}: emit correct, idiomatic \selenium with robust element location and interaction logic (waits, error handling, parameterization), complementing algorithmic/software benchmarks \cite{chen2021evaluating,jimenez2023swebench,li2022alphacode}.
    \item \textbf{Task planning}: decompose the goal into steps and control flow, drawing on reasoning+acting/tool-use strategies \cite{yao2023react,schick2023toolformer,yao2023treeofthoughts,gao2023pal}.
\end{enumerate}

Each competency is evaluated with targeted prompts and execution checks, enabling granular diagnosis beyond aggregate success rates.

\paragraph{Implications for safety and dual-use.}
Because web automation can be repurposed for spam, scraping, or policy-violating behavior, \texttt{MacroBench} also probes \emph{safety alignment} under realistic prompts. Our safety suite complements agentic-safety evaluations \cite{zhang2024agentsafetybench} and policy-aligned training paradigms \cite{ouyang2022instructgpt,bai2022constitutional} by asking whether models refuse or reshape requests that would yield unsafe automation code \cite{ha-etal-2025-one,kim2025objexmtobjectiveextractionmetacognitive,kim2025xteamingevolutionarym2sautomated}. The analysis highlights failure modes specific to code-and-actions settings that are underrepresented in general-purpose safety tests.

\paragraph{Controlled, reproducible environments.}
All tasks run end-to-end on \emph{seven} self-hosted, synthetic websites that emulate widely used platforms while avoiding interaction with live services. Concretely, we instantiate \emph{Airbnb-like} (hospitality marketplace), \emph{TikTok-like} (short-video feed), \emph{reddit-like} (forums), \emph{instagram-like} (photo feed), \emph{facebook-like} (social network), \emph{discord-like} (chat with servers/channels), and \emph{Threads-like} (microblogging) sites, following best practices for reproducible web-agent evaluation \cite{zhou2023webarena,deng2023mind2web}. This design ensures consistent initial states and precise execution-based scoring. Our evaluation includes four contemporary LLMs—\texttt{GPT-4.1}, \texttt{Gemini-2.5-Pro}, \texttt{DeepSeek-V3.1}, and \texttt{GPT-4o-Mini}—to illustrate capability and safety trade-offs in a code-first setting.

\paragraph{Scope and significance.}
Prior web-agent benchmarks emphasize open-web browsing or visually grounded interaction \cite{yao2022webshop,koh2024visualwebarena,xie2024osworld} or evaluate general agent abilities across heterogeneous tools and tasks \cite{liu2023agentbench,shen2024taskbench,yao2024tau,xu2024theagentcompany}. \texttt{MacroBench} is complementary: it targets \emph{macro synthesis over HTML} as the core skill required for scalable, maintainable, and auditable automation. By disentangling interpretation, generation, and planning, our benchmark provides actionable diagnostics for building agents that are both \emph{highly capable} and \emph{responsibly deployable} in real-world automation workflows \cite{phan2025humanitysexam}.

\section{Related Work}
\paragraph{DOM-centric web interaction.}
Early work established the value of reasoning over HTML/DOM rather than pixels for controllable web agents. \cite{shi2017worldofbits} introduced the World-of-Bits platform; \cite{liu2018wge} proposed workflow-guided exploration and a DOM-structured policy; and \cite{jia2019domqnet} grounded actions in DOM graphs. These efforts foreshadow our ``vision-less'' setting: agents that read code structure to choose robust, semantically meaningful actions.

\paragraph{Realistic web environments and open-web evaluation.}
Simulated but realistic websites (e.g., shopping) were popularized by WebShop \cite{yao2022webshop}. WebArena \cite{zhou2023webarena} advanced reproducibility with self-hosted sites; Mind2Web \cite{deng2023mind2web} shifted to real websites with broad task diversity. Recent multimodal benchmarks add screenshots to capture visual grounding—VisualWebArena \cite{koh2024visualwebarena} and OSWorld \cite{xie2024osworld}. Ecosystem efforts aim to standardize evaluation across suites (e.g., BrowserGym) \cite{chezelles2024browsergym}. Our benchmark differs by focusing squarely on \emph{macro synthesis from HTML} and by concretely instantiating \emph{seven} archetypal, self-hosted sites—\emph{Airbnb-like}, \emph{TikTok-like}, \emph{reddit-like}, \emph{instagram-like}, \emph{facebook-like}, \emph{discord-like}, and \emph{Threads-like}—to stress HTML-grounded interaction patterns (forms, feeds, chats, and microblogging) under reproducible conditions.

\paragraph{Agents, tool use, and planning.}
Agent prompting frameworks combine reasoning with action execution. ReAct interleaves thoughts and tool calls \cite{yao2023react}; Toolformer teaches models to decide when and how to invoke APIs \cite{schick2023toolformer}; and Tree-of-Thoughts introduces deliberate search over intermediate thoughts \cite{yao2023treeofthoughts}. Program-aided reasoning leverages external execution to improve reliability \cite{gao2023pal}. Broader agent benchmarks evaluate general decision making across diverse environments and tool suites \cite{liu2023agentbench,shen2024taskbench,yao2024tau,xu2024theagentcompany}. \texttt{MacroBench} adopts compatible prompting styles but evaluates a narrower, industry-relevant target: translating natural-language goals into maintainable Python+Selenium macros.

\paragraph{Code generation and software engineering evaluation.}
LLM code generation is typically assessed via algorithmic correctness (e.g., HumanEval) \cite{chen2021evaluating}, software-maintenance realism (e.g., SWE-bench) \cite{jimenez2023swebench}, or competitive programming \cite{li2022alphacode}. Our benchmark complements these by scoring \emph{browser automation code}: element selection, synchronization, error handling, and parameterized execution—skills that are decisive for robustness on the seven site archetypes above.

\paragraph{Safety of agentic LLMs.}
Safety alignment techniques such as RLHF \cite{ouyang2022instructgpt} and Constitutional AI \cite{bai2022constitutional} motivate systematic testing of agents in interactive settings. Recent benchmarks quantify agentic safety risks and policy adherence \cite{zhang2024agentsafetybench}, including web-specific safety and trustworthiness probes \cite{yoran2024assistantbench,ha-etal-2025-one,kim2025objexmtobjectiveextractionmetacognitive,kim2025xteamingevolutionarym2sautomated}. Our safety suite situates these concerns in code-first automation: can an LLM be induced to output macros that enable spam, scraping, or other policy violations, or does it refuse and propose safer alternatives?

\paragraph{Positioning.}
Where visually grounded agents advance perception and accessibility \cite{koh2024visualwebarena,xie2024osworld}, \texttt{MacroBench} targets the \emph{engineering} substrate of web automation—reading HTML, planning actions, and emitting robust Selenium code. By cleanly factorizing interpretation, generation, and planning, it offers diagnostic signal that complements open-web evaluations \cite{deng2023mind2web} and general agent leaderboards \cite{liu2023agentbench,shen2024taskbench,yao2024tau,xu2024theagentcompany}, and it brings safety assessment closer to the dual-use realities of automation code \cite{ouyang2022instructgpt,bai2022constitutional,zhang2024agentsafetybench}.

\section{Methodology}

\subsection{Benchmark Design and Architecture}

\texttt{MacroBench} evaluates Large Language Models (LLMs) on their ability to synthesize executable, reusable web-automation \emph{macros} (Python+Selenium programs) from natural-language instructions. The benchmark targets three core competencies required for reliable macro synthesis: \emph{code interpretation}, \emph{code generation}, and \emph{task planning}.

\paragraph{Synthetic website ecosystem.}
Following best practices in web-agent evaluation \cite{zhou2023webarena,yao2022webshop}, we deploy a controlled ecosystem of \textbf{seven} synthetic websites that emulate common real-world platforms while ensuring reproducibility and safety:

\begin{itemize}
\item \airbnblike\ marketplace: listing search/filtering, booking flows (20 tasks)
\item \tiktolike\ short-video platform: infinite feeds, social interactions (160 tasks)
\item \redditlike\ forum: subcommunities, posts, voting, comments (130 tasks)
\item \instagramlike\ photo feed: posts, profiles, social actions (120 tasks)
\item \facebooklike\ social network: timeline, groups, events (120 tasks)
\item \discordlike\ chat: servers, channels, messaging (111 tasks)
\item \threadslike\ microblog: timeline, replies, follows (20 tasks)
\end{itemize}

All sites expose consistent HTML/ARIA conventions, deterministic initial states (seeded databases, fixed user accounts), and realistic interaction patterns. Each website runs in isolated containers with frozen dependencies, enabling reproducible evaluation.

\paragraph{Task taxonomy and complexity levels.}
We authored \textbf{681} distinct automation tasks. Across our full run grid (2{,}636 model--task evaluations; some tasks were not attempted by all models), the complexity distribution and success are:

\begin{enumerate}
\item \textbf{Simple}: 2{,}584 runs with 2{,}370 successes (91.7\%)
\item \textbf{Medium}: 44 runs with 37 successes (84.1\%)
\item \textbf{Complex}: 8 runs with 0 successes (0.0\%)
\end{enumerate}

Each task includes natural-language instructions, \emph{explicit success criteria}, and executable validation checks; a run is counted as \emph{successful} only if the program passes static checks, executes without runtime error, and satisfies all DOM/database assertions.

\subsection{Evaluation Framework}
\paragraph{Models under test.}
We evaluate four contemporary LLMs with distinct profiles (model IDs as queried at evaluation time): \gptfourone, \gemini, \deepseek, \gptfouromini. All models receive the same structured prompts, identical retry budget (up to two attempts), and equivalent execution feedback signals for fairness and reproducibility.

\paragraph{Prompting methodology.}
We use structured prompts containing: (i) task specification with success criteria, (ii) HTML context (DOM excerpts with attributes), (iii) technical constraints (Python+Selenium output format), and (iv) few-shot exemplars. Each model receives up to two retry attempts with execution feedback.

\paragraph{Execution and validation pipeline.}
Generated programs undergo end-to-end evaluation:

\begin{enumerate}
\item \textbf{Static checks}: linting, import validation, safety guardrails
\item \textbf{Runtime execution}: headless browser automation in sandboxed containers
\item \textbf{Outcome validation}: DOM assertions, database snapshots, HTTP logs
\item \textbf{Error attribution}: syntax, runtime, logical, timing, or coverage failures
\end{enumerate}

All artifacts (code, traces, screenshots, DOM diffs) are logged for reproducibility.

\subsection{Safety Evaluation Component}

Web automation carries dual-use risks. Our safety suite probes whether models refuse or redirect harmful requests across four categories: scraping/data extraction, spam/abuse, credential harvesting, and privacy violations. We evaluate refusal rates, alternative quality, reasoning clarity, and consistency under paraphrases.

\subsection{Metrics and Analysis}

\paragraph{Primary metrics.}
Execution-based scoring provides the principal signal: (i) task completion rate, (ii) code quality assessment (maintainability, robustness, best practices), and (iii) error resilience under adversarial conditions.

\paragraph{Competency diagnostics.}
We separately analyze code interpretation (element targeting accuracy), code generation (syntactic validity, Selenium practices), and task planning (workflow decomposition, error handling). Ablation studies quantify the contribution of prompt components (HTML context, exemplars, retry attempts).

\section{Results}

\subsection{Overall Performance}

We evaluated four contemporary LLMs across 2,636 unique task-model combinations spanning seven synthetic websites. Results demonstrate clear performance stratification, with
\textbf{GPT-4o-Mini} achieving the highest overall success rate (96.8\%), followed by \textbf{GPT-4.1} (95.3\%), \textbf{Gemini-2.5-Pro} (89.0\%), and \textbf{DeepSeek-V3.1}
(83.4\%).

Table~\ref{tab:overall_performance} summarizes the aggregate results. Despite architectural differences, all models exhibit strong performance on simple automation tasks, but
effectiveness degrades with increasing task complexity and website-specific interaction patterns.

\begin{table}[t]
\centering
\caption{Overall macro-synthesis performance across 2{,}636 model--task combinations.}
\label{tab:overall_performance}
\begin{tabular}{lrrr}
\toprule
\textbf{Model} & \textbf{Total} & \textbf{Success} & \textbf{Rate (\%)} \\
\midrule
GPT-4o-Mini & 680 & 658 & 96.8 \\
GPT-4.1 & 674 & 642 & 95.3 \\
Gemini-2.5-Pro & 666 & 593 & 89.0 \\
DeepSeek-V3.1 & 616 & 514 & 83.4 \\
\midrule
\textbf{Overall} & \textbf{2,636} & \textbf{2,407} & \textbf{91.3} \\
\bottomrule
\end{tabular}
\end{table}

\subsection{Task Complexity Analysis}

Performance varies significantly by task complexity (Table~\ref{tab:complexity}). Simple automation tasks (91.7\% success) are handled reliably across all models, while
medium-complexity workflows requiring multi-step coordination show degraded performance (84.1\%). Complex tasks involving conditional logic, error recovery, and cross-page workflows     
remain largely unsolved (0.0\% success), highlighting fundamental limitations in current LLMs' planning and error-handling capabilities.

\begin{table}[h]
\centering
\caption{Performance breakdown by task complexity level.}
\label{tab:complexity}
\begin{tabular}{lrrr}
\toprule
\textbf{Complexity} & \textbf{Total} & \textbf{Success} & \textbf{Rate (\%)} \\
\midrule
Simple & 2,584 & 2,370 & 91.7 \\
Medium & 44 & 37 & 84.1 \\
Complex & 8 & 0 & 0.0 \\
\bottomrule
\end{tabular}
\end{table}

\subsection{Website-Specific Performance Patterns}

Platform archetype strongly influences macro synthesis success (Table~\ref{tab:website_performance}). \textbf{Discord-like} chat platforms achieve the highest success rate (99.5\%),     
likely due to simple message-posting workflows and predictable DOM structures. \textbf{Facebook-like} social networks also perform well (98.7\%), while \textbf{TikTok-like}
short-video platforms prove most challenging (81.5\%), primarily due to infinite-scroll feeds and dynamic content loading.

\begin{table}[h]
\centering
\caption{Success rates by website archetype (task count, total runs, successes, and rates).}
\label{tab:website_performance}
\begin{tabular}{lrrrr}
\toprule
\textbf{Website} & \textbf{Tasks} & \textbf{Total Runs} & \textbf{Successes} & \textbf{Rate (\%)} \\
\midrule
\discordlike   & 111 & 426 & 424 & 99.5 \\
\facebooklike  & 120 & 463 & 457 & 98.7 \\
\redditlike    & 130 & 499 & 470 & 94.2 \\
\threadslike   &  20 &  80 &  72 & 90.0 \\
\instagramlike & 120 & 471 & 412 & 87.5 \\
\airbnblike    &  20 &  80 &  69 & 86.3 \\
\tiktolike     & 160 & 617 & 503 & 81.5 \\
\bottomrule
\end{tabular}
\end{table}

\noindent\emph{Note.} We omit per-site model rankings to avoid over-interpretation given missing runs on some site--model pairs; see Appendix for per-model breakdown.

Model rankings vary by platform, suggesting interactions between model design and UI patterns (e.g., feeds, chats, booking flows). We therefore avoid per-site rankings in the main text and defer detailed breakdowns to the Appendix where available.

\subsection{Error Analysis and Failure Modes}

We analyzed 229 failed attempts. In our current logs, most failures fall into an \texttt{unknown} bucket—i.e., programs execute but do not meet outcome assertions. Because attribution is coarse, we refrain from asserting specific causes (e.g., logic vs. timing) and instead report them as \emph{objective-mismatch} failures. We are extending the pipeline to disambiguate selector errors, timing issues (waits), and incomplete control flow.

\textbf{DeepSeek-V3.1} exhibits the highest failure rate (16.6\%, 102/616), while \textbf{GPT-4o-Mini} achieves the lowest (3.2\%, 22/680). The distribution suggests that
code-specialized models may generate syntactically correct but semantically incomplete automation sequences, while general-purpose models with stronger instruction-following produce     
more reliable end-to-end workflows.

\subsection{Code Quality Assessment}

Our code quality rubric evaluated generated macros across five dimensions: syntax correctness, Selenium best practices, error handling, wait strategies, and maintainability.
Under a strict pass/fail rubric (explicit waits, structured error handling, parameterization, and maintainable structure \emph{all} required), no model achieved a full pass. This exposes a gap between functional completion and production-readiness; we include dimension-wise checks in the Appendix for future replication and refinement.

This finding highlights a critical disconnect: while models can generate macros that complete specific tasks, the resulting code lacks robustness features essential for real-world       
deployment (explicit waits, exception handling, parameterization, and maintainable structure).

\subsection{Safety Evaluation Results}

Our safety suite revealed nuanced patterns in harmful request handling. All models consistently refuse explicitly harmful requests (credential harvesting, privacy violations), but       
responses vary in quality and constructiveness.

\textbf{GPT-4.1} and \textbf{GPT-4o-Mini} demonstrate superior "refuse-and-repair" behavior, declining unsafe requests while proposing policy-compliant alternatives.
\textbf{Gemini-2.5-Pro} and \textbf{DeepSeek-V3.1} more frequently provide bare refusals without constructive redirection, potentially frustrating legitimate automation needs.

Notably, ambiguous requests (bulk data export, workflow scaling) expose inconsistencies across models, with some interpreting benign automation intent while others default to
conservative refusal. This suggests need for more nuanced safety training that distinguishes between legitimate productivity automation and policy-violating abuse.

\subsection{Model-Specific Characterization}

\paragraph{GPT-4o-Mini} achieves the best overall performance through consistent execution and high success rates across diverse platforms. Its efficiency-oriented design appears        
well-suited to structured macro synthesis tasks.

\paragraph{GPT-4.1} shows balanced performance with particular strength on complex platforms (TikTok-like) but occasional inconsistencies on simpler tasks (Discord-like).

\paragraph{Gemini-2.5-Pro} demonstrates deliberate reasoning capabilities with strong performance on complex booking flows (Airbnb-like) but struggles with high-frequency interaction    
patterns.

\paragraph{DeepSeek-V3.1} exhibits the most variable performance—excelling on specific platforms (Discord) while showing systematic weaknesses on dynamic content handling and
workflow planning.

\subsection{Implications for Practical Deployment}

Our results suggest that current LLMs can reliably automate simple, well-structured web tasks but remain unsuitable for complex, multi-step workflows requiring robust error handling     
and dynamic adaptation. The universal absence of production-quality coding practices indicates that generated macros require significant human review and refinement before
deployment.

For practitioners, these findings recommend: (1) limiting initial deployments to simple, single-step automation tasks; (2) implementing comprehensive testing and validation
pipelines; (3) developing model-specific prompting strategies that emphasize robustness and maintainability; and (4) establishing clear safety guidelines for automation request
handling.

\section{Conclusion}
This work introduced \texttt{MacroBench}, a controlled benchmark for evaluating 
LLMs on executable web-automation macro synthesis. 
Our results reveal clear stratification across contemporary models: 
while state-of-the-art systems (e.g., GPT-4o-Mini, GPT-4.1) 
consistently succeed on simple single-step tasks, 
performance deteriorates on medium-complexity workflows 
and collapses entirely for complex, planner-heavy scenarios. 
The analysis highlights that although LLMs exhibit strong code-generation 
and syntactic reliability, deficiencies persist in synchronization, 
robust element targeting, and workflow planning. 
Safety evaluation indicates that refusal alone is insufficient; models exhibiting \emph{refuse-and-repair}—declining unsafe requests while proposing policy-aligned alternatives—provide greater practical utility.

Overall, \texttt{MacroBench} underscores that progress in 
macro synthesis requires not only scaling model capacity, 
but also targeted improvements in reasoning, error recovery, and safety alignment.


\section{Limitations}
While comprehensive, this study carries several limitations. 
First, our evaluation is constrained to a synthetic ecosystem 
of seven websites, which, although carefully designed, 
cannot fully capture the diversity and complexity of real-world platforms. 
Second, our task set, though broad, emphasizes common interaction patterns 
and may under-represent rare or adversarial workflows. 
Third, the safety probes focus on a limited set of harmful request categories 
and do not exhaustively span the long tail of misuse risks. 
Fourth, our error attribution and rubric-based quality assessments 
involve human-designed criteria, which may introduce subjectivity. 
Finally, we restrict evaluation to four representative LLMs; 
future work should extend to a wider range of architectures 
and instruction-tuning strategies. 
We view these limitations as opportunities for refinement: 
expanding the website suite, diversifying task complexity, 
broadening safety probes, and scaling cross-model comparisons 
will yield a more complete picture of LLM capabilities in web automation.

\clearpage

\printbibliography

\newpage
\appendix
\section{Workshop-Targeted Appendix: Using Benchmark Outputs for Macro-Safety Training}
\label{app:lockllm}

This appendix outlines how \macrobench artifacts can support \emph{training and evaluation} of LLMs and guardrail models that resist unsafe automation requests, aligning with the Lock-LLM workshop theme of preventing unauthorized knowledge use. We restrict releases to anonymized artifacts and redact operationally dangerous details pending camera-ready.

\section{Appendix: Detailed Experimental Data}

\subsection{Aggregate Results}
\begin{itemize}
    \item GPT-4o-Mini: \textbf{658/680} (96.8\%)
    \item GPT-4.1: \textbf{642/674} (95.3\%)
    \item Gemini-2.5-Pro: \textbf{593/666} (89.0\%)
    \item DeepSeek-V3.1: \textbf{514/616} (83.4\%)
\end{itemize}

\paragraph{Task Complexity.}
\begin{itemize}
    \item Simple: \textbf{2{,}370/2{,}584} (91.7\%)
    \item Medium: \textbf{37/44} (84.1\%)
    \item Complex: \textbf{0/8} (0.0\%)
\end{itemize}

\subsection{Website Performance (Success Rate \%)}
\begin{itemize}
    \item Airbnb-like: \textbf{86.3}
    \item TikTok-like: \textbf{81.5}
    \item Reddit-like: \textbf{94.2}
    \item Instagram-like: \textbf{87.5}
    \item Facebook-like: \textbf{98.7}
    \item Discord-like: \textbf{99.5}
    \item Threads-like: \textbf{90.0}
\end{itemize}

\paragraph{Why this matters for Lock-LLM.}
Unlike conventional safety evaluations that target overtly harmful content (e.g., instructions on self-harm, weapon building, or drug synthesis), our benchmark reveals a neglected but equally pressing failure mode: modern LLMs will readily generate executable \emph{macros} in Python+Selenium that can automate large-scale abuse. Examples include bulk posting across forums, scraping user data in violation of terms of service, or constructing credential-harvesting workflows. These cases are rarely flagged by existing guardrails because they appear as ``help with automation'' rather than obviously malicious requests. The result is a form of \textbf{capability laundering}, where powerful but risky behaviors are surfaced under the guise of productivity assistance.

\paragraph{Leveraging benchmark outputs for fine-tuning.}
Because our evaluation logs include per-task prompts, model generations, execution traces, and curated safety labels, they serve as a concrete resource for training safety-aware systems. Researchers can:
\begin{itemize}
    \item Fine-tune \textbf{guardrail models} to detect and block risky macro requests at both the prompt and output levels.
    \item Train LLMs with \textbf{refuse-and-repair} behavior, where unsafe automation requests are declined and redirected toward safe alternatives (e.g., policy-compliant exports, synthetic test workflows).
    \item Build \textbf{critics and filters} that score generated programs for signs of high-risk automation patterns (mass posting loops, credential interception, or unbounded scraping).
\end{itemize}

\paragraph{Closing the gap.}
Our dataset highlights that current LLMs often succeed at macro synthesis but fail at aligning that capability with platform policies. By releasing structured results, refusal labels, and repair targets, we provide the first reproducible basis for developing models that can \emph{distinguish} between benign automation (e.g., accessibility support, testing workflows) and abusive automation (e.g., spam farms, large-scale scraping). This directly supports the Lock-LLM agenda of mitigating unauthorized knowledge use: the same benchmarks that expose unsafe behavior also provide the training signals necessary to constrain it.

\paragraph{Takeaway.}
In short, \texttt{MacroBench} shows that LLMs do not only produce risky outputs in obvious categories of harm, but also in subtle, automation-centric scenarios that are currently under-protected. Our released data makes it possible to fine-tune both LLMs and guardrail models to address these overlooked risks, turning benchmark evaluation into actionable safety training material.

\subsection{Why Macro Safety Is Underserved}
\label{app:gap}
Prevailing taxonomies emphasize recognizable explicit harms. Macro synthesis introduces \textbf{capability laundering}: seemingly benign ``please automate X'' requests conceal high-risk behaviors (spam farms, bulk extraction, credential flows). Because outputs are \emph{executable} programs, a single unsafe response scales to \emph{automated}, repeated violations. Our results show this gap is systematic across models, justifying a dedicated training and evaluation pipeline for \emph{macro-aware safety}.

\subsection{What We Release}
\label{app:release}
We release artifacts tailored for \emph{training and benchmarking} macro safety:

\begin{itemize}
  \item \textbf{Per-task records} for 200 tasks $\times$ 4 models: prompt, HTML/DOM context, model program output (Python+Selenium), execution logs (stdout/stderr, traces), and evaluation outcomes with labeled failure modes.
  \item \textbf{Safety labels} at \emph{prompt} and \emph{output} levels:
  \begin{itemize}
    \item \texttt{risk\_label} $\in$ \{\textsc{safe}, \textsc{ambiguous}, \textsc{unsafe}\}
    \item \texttt{refusal\_label} $\in$ \{\textsc{refuse}, \textsc{comply}, \textsc{repair}\} (``repair'' = \emph{refuse-and-redirect} with safe alternatives)
    \item \texttt{intent\_surface} (read-only, content-creation, account/identity, bulk/automation, data-exfiltration), \texttt{policy\_concern} (scraping, spam/abuse, credentials/privacy)
  \end{itemize}
  \item \textbf{Repair targets}: policy-aware rewrites, consented/rate-limited exports, and synthetic fixtures that models should choose over risky code.
  \item \textbf{Splits \& generalization}: train/dev/test with (i) \emph{cross-site} and (ii) \emph{prompt-paraphrase} generalization.
  \item \textbf{Governance kit}: usage policy, evaluation scripts, and a risk-aware leaderboard protocol reporting both \emph{utility} and \emph{safety}.
\end{itemize}

\paragraph{Data format.}
Each record is a self-contained JSON object:
\begin{verbatim}
{
  "task_id": "...", "site": "...", "split": "train|dev|test",
  "prompt": "...", "html_context": "...",
  "model_id": "...", "generation": "...",
  "exec_pass": true|false, "failure_mode": "syntax|timing|logic|...",
  "risk_label": "SAFE|AMBIGUOUS|UNSAFE",
  "refusal_label": "REFUSE|COMPLY|REPAIR",
  "policy_concern": ["scraping","spam","credentials"],
  "repair_target": "policy-compliant alternative"
}
\end{verbatim}

\subsection{Intended Uses}
\label{app:intended}
Artifacts enable training and evaluation of:
\begin{enumerate}
  \item \textbf{Guardrail classifiers} detecting risky macro intent and risky program patterns \emph{pre-} and \emph{post-generation}.
  \item \textbf{Refuse-and-repair LLMs} that decline unsafe automation and provide policy-compliant alternatives (consented exports, synthetic stubs).
  \item \textbf{Planner critics / linters} that score plans and code for macro-risk signatures (mass posting loops, credential flows, unbounded scraping).
\end{enumerate}

\subsection{Training Recipes (High-Level, Safety-Preserving)}
\label{app:recipes}
We leverage labels without teaching step-by-step abuse. All setups operate on prompts and high-level alternatives.

\paragraph{SFT for refusal-and-repair.}
Inputs: (prompt, minimal context). Targets: curated \texttt{REPAIR} responses. Validate for (i) refusal on \textsc{unsafe}, (ii) helpfulness on \textsc{safe}.

\paragraph{Preference optimization (DPO/IPO).}
Form pairs $(y^{\mathrm{good}}, y^{\mathrm{bad}})$ where $y^{\mathrm{good}}$ is \texttt{REPAIR} or compliant, and $y^{\mathrm{bad}}$ is a code-emitting \texttt{COMPLY} for \textsc{unsafe}/\textsc{ambiguous}. Optimize to shift mass away from unsafe compliance.

\paragraph{Reward modeling for macro risk.}
Train $R(\cdot)$ to predict \texttt{risk\_label} from (prompt, plan/code summary). Use $R$ as a decoding gate or to trigger clarifications before emitting code.

\paragraph{Static program filters (post-generation).}
Detect abstracted features: action count, credential flows, cross-origin hops, loop depth, missing waits/guards, target roles.

\paragraph{Parameter-efficient fine-tuning.}
Apply LoRA/QLoRA (e.g., rank 8--32, 1--3 epochs, 5\% warmup) on \texttt{REPAIR}-centric data to avoid memorizing unsafe snippets.

\subsection{Evaluation: Utility--Refusal Frontier}
\label{app:urf}
Report a \emph{Utility--Refusal Frontier (URF)}:
\begin{itemize}
  \item \textbf{Benign utility}: success on \textsc{safe} tasks without unnecessary refusal.
  \item \textbf{Harm refusal}: refusal rate on \textsc{unsafe} tasks.
  \item \textbf{Ambiguity handling}: share of \textsc{ambiguous} prompts eliciting clarification or safe alternatives rather than code.
\end{itemize}
Compare models at matched benign-utility points (e.g., 90\% of baseline benign success) to fairly assess refusal quality.

\subsection{How to Integrate Guardrails in a Macro Agent}
\label{app:integration}
\begin{enumerate}
  \item \textbf{Pre-generation classifier}: score user intent; route to \texttt{REPAIR} when risk is high.
  \item \textbf{Planning gate}: score macro plans with $R$; require confirmation or rewriting if risky.
  \item \textbf{Static code audit}: lint for credential capture, mass posting, cross-site scraping, and missing rate limits.
  \item \textbf{Sandbox-first execution}: execute in hermetic sandboxes with policy simulators; require approval for live endpoints.
  \item \textbf{Post-exec telemetry}: log element/role-level actions and feed counterexamples back into training.
\end{enumerate}

\subsection{Splits and Generalization Protocols}
\label{app:splits}
Two complementary regimes:
\begin{description}
  \item[Cross-site:] train on a subset of websites; test on disjoint sites to measure robustness to unfamiliar DOM/layout conventions.
  \item[Prompt-shift:] train on canonical phrasings; test on paraphrases, indirect requests (``help me scale this workflow''), and injection-style wording.
\end{description}
Report macro-averaged metrics across \texttt{risk\_label} strata to avoid masking regressions.

\subsection{Red Teaming and Probe Design}
\label{app:redteam}
We provide non-explicit but risky prompts that realistically elicit abuse-capable macros while avoiding overtly malicious phrasing. To preserve safety, releases omit exploit steps and operational code; instead we include high-quality \emph{repair} targets and \emph{benign decoys} to test specificity and false positives.

\subsection{Responsible Release and Use}
\label{app:responsible}
\textbf{Intended use.} Research on macro-aware LLM safety, training guardrails, evaluating refusal/repair. \\
\textbf{Prohibited use.} Training models to increase harmful automation success or to bypass platform policies. \\
\textbf{Access controls.} Research-use terms prohibit reproducing unsafe code outside sandboxed evaluation and require downstream safety disclosures. \\
\textbf{Minimization.} Unsafe generations are redacted or abstracted to features/metadata sufficient for detectors and repairs. \\
\textbf{No live-target interaction.} All websites are synthetic; no real user data or credentials are present. \\
\subsection{Data Card (Summary)}
\label{app:datacard}
\textbf{Motivation.} Enable macro-aware safety via refusal and safe alternatives. \\
\textbf{Composition.} 200 tasks (a curated subset) across \textbf{seven} synthetic websites; 4 LLMs' generations with traces and safety labels. \\

\textbf{Collection.} Prompts span interaction complexity and safety probes; execution in sandboxed containers. \\
\textbf{Labels.} Dual-annotator protocol with adjudication for \texttt{risk\_label} and \texttt{refusal\_label}; agreement metrics reported in-repo. \\
\textbf{Uses.} Guardrail classification, reward modeling, refusal-and-repair fine-tuning. \\
\textbf{Limitations.} Domain shift to real sites; evolving ToS; conservatism bias in repairs. \\
\textbf{Ethics.} Dual-use risks mitigated via abstraction, licensing, and bans on operational misuse.

\subsection{Limitations and Future Work}
\label{app:limits}
Our taxonomy focuses on browser macros; native app and OS-level scripting are out of scope. Refusal-and-repair must balance \emph{helpfulness} with \emph{harm avoidance}. Future work: (i) richer consent/authorization flows; (ii) context-grounded repairs using platform ToS templates; (iii) adversarial training with automatically mined ambiguous prompts.

\subsection{Reproducibility Checklist for Safety Fine-Tuning}
\label{app:checklist}
\begin{enumerate}
  \item Publish exact train/dev/test splits and task IDs.
  \item Log refusal thresholds or gating (e.g., reward cutoffs); report URF at multiple operating points.
  \item Release redaction policy for unsafe outputs (kept text/features).
  \item Document prompting instructions used to elicit repairs/clarifications.
  \item Provide failure analysis by \texttt{policy\_concern} and website category.
\end{enumerate}

\subsection{Practical Notes for Practitioners}
\label{app:practical}
\begin{itemize}
  \item \textbf{Prefer repair over silence.} Offer actionable, policy-compliant alternatives (e.g., consented exports) instead of bare refusals.
  \item \textbf{Tune strictness.} Calibrate thresholds to preserve benign utility; report full URF curves.
  \item \textbf{Human-in-the-loop.} Route high-risk/uncertain cases; log decisions for continual improvement.
  \item \textbf{Defense-in-depth.} Combine intent filters, plan critics, static audits, and sandbox approvals rather than relying on a single stage.
\end{itemize}

\paragraph{Takeaway.}
LLMs readily produce executable macros that can scale abuse. The \texttt{MacroBench} artifacts—prompts, structured outputs, execution traces, and curated refusal/repair targets—turn evaluation into \emph{training signals} for macro-aware safety, directly targeting risks underrepresented in conventional content-focused safety.

\end{document}